\documentclass[draft,12pt]{article}
\usepackage{amsmath,amsfonts,palatino,amsthm,epsf}
\theoremstyle{definition}

\newcommand{\beq}{\begin{equation}}
\newcommand{\eeq}{\end{equation}}

\newcommand{\f}{\begin{equation}}
\newcommand{\ff}{\end{equation}}
\newcommand{\AL}{{\rm AL}}

\begin{document}
\title{Gauge fixing \\
 in Causal Dynamical Triangulations}
\author{
Fotini Markopoulou\thanks{Email address: 
fotini@perimeterinstitute.ca}\ \ and 
Lee Smolin\thanks{Email address:lsmolin@perimeterinstitute.ca}\\ \\
Perimeter Institute for Theoretical Physics,\\
35 King Street North, Waterloo, Ontario N2J 2W9, Canada, and \\
Department of Physics, University of Waterloo,\\
Waterloo, Ontario N2L 3G1, Canada\\
\\
}
\date{\today}
\maketitle

\begin{abstract}
We relax the definition of the Ambj{\o}rn-Loll causal dynamical triangulation model in $1+1$ dimensions to allow
for a varying lapse.  We show that, as long as the spatially averaged lapse is constant in time, the physical observables are
unchanged in the continuum limit.   This supports the claim that the time slicing of the model is the result of a gauge
fixing, rather than a physical preferred time slicing.

\end{abstract}
\vfill

\section{Introduction}

A widely accepted strategy to construct a quantum theory of gravity is to define the theory by means of a background independent path-integral, in which one sums over microscopic spacetime histories, each weighed by a quantum amplitude.  Such a theory can be a candidate for a fundamental theory of spacetime because it takes over into the quantum realm the most basic, and experimentally confirmed, principle of general relativity, namely, that the geometry of space and time is entirely dynamical.  Approaches of this kind include causal sets, loop quantum gravity, spin foam models, Regge calculus and dynamical triangulations. All path-integral approaches have the same ultimate aim,  to
derive general relativity as a low energy, coarse grained 
approximation to the fundamental discrete and quantum description.  
This is still an open problem and we will call it  {\it the low energy problem.}

In almost all cases, the sum-over-histories is a formal path-integral.  At the mathematical level, that path integral suffers from being ill-defined, and  in several ways.  At the practical level,  it is very complicated, making the low energy problem a formidable one.
It is no less obscure at the physical level, where its formal interpretation as quantum superpositions of spacetimes needs to be understood and related to observations. 

The overall motivation for the present work is to investigate a basic questions raised by sum-over-histories quantum gravity:  {\it which, if any, geometric features are shared between the quantum histories,  ``the paths'',  and the classical spacetime geometry that is to emerge in the low energy limit?}  In particular, what is the relationship between the fundamental microscopic causal structure and the macroscopic causal structure that emerges in the low energy limit? Similarly, what is the
relationship between microscopic and macroscopic locality?

These questions can only be investigated in the context of  a path-integral model where explicit results on the low energy limit can be calculated. 
Of all the different approaches, the one that has been most successful
at addressing the low-energy  problem is 
Causal Dynamical Triangulations (CDT).  This is a non-perturbative gravitational path-integral
for Lorentzian spacetimes, proposed by Ambj{\o}rn and Loll in \cite{AL}.  The  
histories are described, in the presence of an 
ultraviolet regulator, $a_s$, as a triangulated Lorentzian manifold.
Each spacetime history is built by pasting together fundamental 
building blocks with timelike and
spacelike faces whose interiors are isomorphic to subsets of Minkowski
spacetime.  Each history has a causal structure,  which means that the histories are restricted to contain arrangements of simplices whose spacelike faces form a sequence of spatial slices, the triangulated equivalent of a globally hyperbolic spacetime with equal-time slices.  

In $1+1$ dimensions, Causal Dynamical Triangulations have been shown, 
both analytically and numerically, to have a continuum limit when
$a_s \rightarrow 0$ which describes physics in an effective $1+1$ 
geometry, which is large compared to the Planck scale. One 
infers this from measurements of the Hausdorff dimension, which is near
$2$ (instead of the problematic  $4$ of Euclidean dynamical triangulations \cite{dynamical}).
The good behavior is maintained when matter is added, and it 
is even possible to exceed the barrier of $c=1$ that prevented the 
Euclidean versions of such models from providing a 
non-perturbative approach to string theory\cite{c=1}. 
Recently published numerical results in $3+1$ dimensions provide 
evidence for a continuum limit which consists of universes which
are $3+1$ dimensional, whose spatial slices have volumes which are
large and slowly varying, and whose spatial geometries have
Hausdorff dimension $d_h \approx 3$ \cite{AL3+1}.

Compared to other path-integral approaches, Causal Dynamical Triangulations have three distinct features: 

\begin{enumerate}

\item{}Each history is a fixed foliation of spacelike
surfaces and thus has topology $\Sigma\times R$, with $\Sigma$ a fixed spatial topology.

\item{}Since  each triangle has fixed spatial edge length $a_s$ and fixed timelike edge length $a_t$, the fixed foliation implies that the lapses are fixed to be a constant, $N_0=\sqrt{a^2_t-\frac{a_s^2}{4}}$,  in the $1+1$ dimensional model.  The same is true  in higher dimensions.  

\item{}The limit $a_s \propto a_t \rightarrow 0$ is taken,
so that the discreteness scale is considered an ultraviolet cutoff  to be removed rather than a fixed physical scale.  

\end{enumerate}

To understand the results of CDT models, and compare them to results that may be expected from other approaches, it is essential to know the extent to which these three features are necessary conditions for the good results obtained.  
In particular, we would like to know to what extend the CDT histories have a geometrical interpretation.  The model can be understood as  a statistical model of random surfaces, and the inventors of Causal Dynamical Triangulations warn their readers not to think of the individual histories as spacetimes.  However, if we strictly follow their advice, then in what sense should the constraints they impose on the random surfaces that are allowed to enter the sum be interpreted as causality?  Can different causality conditions be imposed that lead to the same  results?

The specific task for this paper is the following:
Given a global foliation, and hence a $\Sigma \times R$ topology, 
the choice of equal lapses can be conjectured to be a gauge
condition. If this is the case it should be possible to show
explicitly that the results of CDT are
preserved if the gauge conditions is modified.

We will work in the $1+1$ case, and show that the same continuum limit results are obtained if the lapses of the simplices are allowed to vary over positive values, subject only  to two conditions.  1)   the average lapse in each time slice is fixed.  
2)  two edges of each triangle remain timelike.
This applies both to the pure gravity model and to gravity coupled to matter.  This supports the claim that, once the $\Sigma\times R$ topology is fixed, the condition of fixed lapses is just a gauge condition that can be modified without affecting the physical results.  

In a related paper\cite{tomasz},  Konopka shows that similar results
can be obtained for the $2+1$ case.  Together with the results presented here, 
this gives strong support for the claim that the restriction to constant lapse
in CDT models can be considered as a gauge condition.  

One of the main issues with the preferred foliation of this model is the implication of non-locality in the causal evolution.  The rule that constructs the histories is non-local because the spacelike slices that define the preferred foliation are first fixed, after which one sums over all triangulations that connect them, subject to the condition that two edges of each triangle remain timelike.  This can
be contrasted with other causal approaches to the path integral in quantum gravity, 
such as causal spin networks \cite{CSN} or percolation causal sets \cite{pcausalset}.  There, histories are constructed by {\it local} evolution rules, meaning that the  rule for adding a triangle and computing its amplitude depends only on its causal past.  The existence of the  global foliation imposed in CDT models requires either the application of a non-local rule, or the truncation of the ensemble of histories constructed by local rules.   

Our results indicate that the non-locality is still present even when the lapse gauge is changed, since it is required to keep the average lapse constant.  The class of histories with local causal rules is much larger than the Causal Dynamical Triangulations, and there is little reason at this point to believe that the good  CDT results are compatible with local causal rules.   It appears that, while other features of sum-over-histories quantum gravity, such as locality or fundamental discreteness, appear desirable, they are not shared by Causal Dynamical Triangulations.  As it is the latter that possesses the good low-energy behavior, this implies either that these desirable features are misguided, or that our previous expectations for the geometrical interpretation of the path-integral histories as the  microscopic description of spacetime were naive. 

The outline of this paper is as follows:  In Section 2 we review the 1+1 Causal Dynamical Triangulations.  In Section 3, we generalize the 1+1 pure gravity model to the varying lapse gauge.  In Section 4, we apply the varying lapse gauge to gravity coupled to matter.   In Section 5 we compare Causal Dynamical Triangulations to local causal models.  We discuss our conclusions in Section 6.

\section{Pure gravity 1+1 Causal Dynamical Triangulations}

Causal Dynamical Triangulations is a non-perturbative gravitational 
path integral for Lorentzian spacetimes, proposed by Ambj{\o}rn and Loll 
in \cite{AL}.   Computing the path integral is done by counting geometries
at the regularized level weighted by the Lorentzian Einstein action.  The 
regularized spacetimes are histories of flat Lorentzian simplices with certain 
causality conditions.  

In pure gravity in $1+1$ dimensions, the only attribute a spatial
slice has is its length $L$.  In the regulated model $L=a_s l$, where $a_s$ is the
spacelike edge length, which serves as an ultraviolet regulator in the
model, and $l$ is the number of spacelike edges that make up the slice, 
an integer. 

The central object in the Ambj{\o}rn-Loll pure gravity model is
the amplitude to go from a universe of length $l_0 $ to a universe
with length $l_t $ in $t $ time steps:
\f
G(l_0, l_t, t) = \sum_{ T(t):l_0\rightarrow l_t} e^{i S_{\AL}[ T(t) ]}
\ff
where $T(t)$ refers  to all the triangulations
with in boundary a circle of length $l_0$, out boundary of length $l_t$ and 
consist of $t$ time steps (having chosen periodic boundary conditions).  

A history with $t$ time steps has $t+1$ time slices $i=0,...,t$,
each of length $l_i > 0$.  Between slice $i$ and $i+1$ is a band of 
$n_{i+1}= l_i + l_{i+1}$ triangles, with $l_i$ triangles oriented ``up"
with base on the bottom slice and $l_{i+1}$ triangles oriented down,
with base on the top slice:
\[
\begin{array}{c}\mbox{\epsfbox{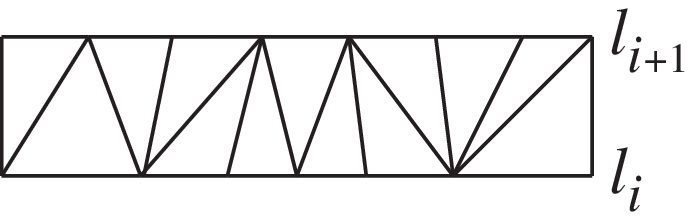}}\end{array}
\]
Each triangle has one spacelike edge of length $l_s^2=a^2$ and two equal 
time like edges of length $l_t=-a^2$ where $a$ is the regularization cutoff:
\[
\begin{array}{c}\mbox{\epsfbox{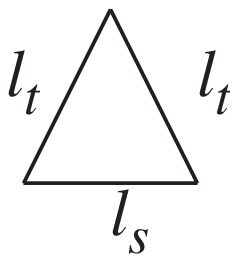}}\end{array}
\]
The height of the triangles is $\Delta t=1$.
Note that the causality conditions of the model is not restricted to the Minkowskian 
interior of the triangles, but also require that there is no spatial topology change or closed timelike curves.  

In two dimensions, the Einstein action is a topological invariant, 
\beq
\int_M dx \sqrt{|\det g|}R=2\pi\chi,
\eeq
where $\chi$ is the Euler characteristic of the spacetime $M$.   In the CDT model, 
only trivial topologies are allowed so weighing the histories by this action simply 
gives a constant overall factor to the path integral.  Since in a dynamical triangulation scheme each triangle has the same
area, the action for each history is
\f
S_{\AL}[T]= \Lambda  A n(T),
\ff
where $n(T)$ is the number of triangles in the history.
The area is defined by $A= N_0 a_s /2$ where $N_0$ the fixed lapse.  

Note that it is part of the definition that
the length of the universe can change by any amount in a single time
step (but cannot shrink to less than $l=1$). 
We also note that the physical amplitude should be divided
by $l_0$ to divide out by a discrete residue of spatial
diffeomorphism invariance at the boundaries. (i.e. histories that
differ by an overall rotation of the initial universe are
equivalent.)

The propagation amplitude $G(l_0, l_t, t)$ satisfies
\begin{equation}
G(l_0, l_t, t)= \sum_{l=1}^{\infty} G(l_0, l , 1) G(l, l_t, t-1)
\label{eq:G}
\end{equation}
where the single time step amplitude $ G(l_0, l , 1)$ is the 
one-step transfer matrix. It can be written as follows.
Each vertex $r$ of the initial ring is attached
to $k_r$ vertices of the final ring, which we call the
future set of the vertex $r$. However we 
identify the leftmost vertex in the future set of $r$
with the rightmost vertex in the future set of $r-1$.  There
are then $n_0= l_0 +l_1 = \sum_{r=1}^{l_0} k_r$ triangles
in the strip, and hence the one-step transfer matrix is 
\f
G(l_0, l , 1)= \sum_{k_1,..., k_{l_0}} 
e^{i \Lambda  A\sum_{r=1}^{l_0} k_r}.
\ff
It is convenient to define the amplitude for a single triangle by 
\f
g=e^{i \Lambda A},
\ff
in which case the amplitude of a history is $g^{n(T)}$.

To solve eq.\ (\ref{eq:G}) we invent the generating function
\f
\tilde{G}(x,y,t) = \sum_{k,l=1}^{\infty} x^k y^l G(k,l,t),
\label{generate}
\ff
so that the recursion relation (\ref{eq:G}) is written for the 
generating function as
\f
\tilde{G}(x,y,t_1 + t_2)= \oint {dz \over 2\pi i z} 
\tilde{G}(x,z^{-1},t_1) \tilde{G}(z,y,t_2).
\ff

One then arrives at the transfer matrix 
$\tilde{G}(x,y,1)$ for the transform by a counting argument.
We use (\ref{generate}) to construct the generating
function for one time step, which is the amplitude to
go from an initial loop to a next loop. We sum over the number of
edges in the initial loop, and associate a factor of $x$ with each
edge, coming from  the definition (\ref{generate}). Each
initial edge can give rise to any number of edges in the next
loop.  To each of these we associate a factor $y$.  We then multiply
each triangle by a factor $g$. There is one triangle (pointed up)
for each edge in the initial loop, and one triangle, (pointed down)
for each triangle in the next loop.  Thus each $x$ and each $y$ is
multiplied by a factor of $g$.  The amplitude is given by
the double sum
\f
\tilde{G}(x,y,1)= \sum_{k=0}^\infty \left (
gx \sum_{l=0}^\infty (gy)^l
\right )^k - \sum_{k=0}^\infty (gx)^k = \frac{g^2 xy}{(1-gx)(1-gx-gy)}.
\label{eq:gen2}
\ff
The second factor excludes the possibility that the next loop has zero 
edges.

The analysis proceeds from here by the solution of the
recursion relation
\f
\tilde{G}(x,y,t )= \oint {dz \over 2\pi i z} 
\tilde{G}(x,z^{-1},1) \tilde{G}(z,y,t-1) . 
\label{recursion}
\ff
This leads to the construction of the full
generating function, which is then a function only of $x,y$ and $g$ (for more details see \cite{AL}).
It is then shown that the continuum limit is possible because of the
existence of a fixed point at 
\f
g_c=\pm \frac{1}{2}.
\label{eq:gcrit}
\ff
If we define the dimensionless quantity
\f
\lambda = \Lambda A,
\ff
the critical point (\ref{eq:gcrit}) corresponds to a critical $\lambda_0=i \ln 2$.

\section{Pure gravity CDT in varying lapse gauge}

 In the
quantization of a diffeomorphism invariant theory the lapses must be
allowed to vary without affecting the physical results. If that is the case in CDT, then 
the fixing of the lapses is merely a gauge
condition and not a physical reduction of the degrees of freedom of 
the theory.  To examine whether this is the case, 
we now
 consider a simple generalization of CDT in which the lapses and areas of the 
different triangles in a triangulation are allowed to vary.  

We denote $N_i$ the lapse of triangle $i$.  If we keep the spatial edge lengths $a_s$ fixed for all triangles, corresponds to area $A_i=\frac{1}{2} N_i a_s$.  We require that 
\f
N_i > N_{\rm min}= \frac{a_s}{2}
\ff
below which the  the timelike edges in the triangles become null, then spacelike. 

With this change the action for a history $T$ becomes
\f
S[T]=\Lambda\sum_{i\in T}A_i=\Lambda\bar{A}n(T)
\ff
where $\bar{A}=\langle A_i\rangle$ is the average triangle area and $n(T)$ is the number of triangles in triangulation $T$.   The new weight for a triangle becomes
\f
g_i=e^{i\Lambda A_i}.
\ff

The CDT model is solved using a one-step transfer function.  To reproduce this solution for the varying lapses, one has to work with average lapse and area in a single slice, instead of the average history lapse $\bar{N}=\langle N_i\rangle$.  A ``spacetime'' solution of the 1+1 CDT that would accommodate the average history lapse is underway elsewhere \cite{AM}.   For the present note, we will be restricted to a generalization of the CDT model that is easily constructed within the present transfer matrix solution.  For this, we need to restrict to histories in which the average lapse of each slice {\it remains constant in time}: 
\f
\bar{N_t}=\bar{N}.
\ff

At this point, it is clear that our modest generalization cannot change the CDT partition function. The action changes
by an overall scaling.
\f
S[T]=\Lambda \frac{\bar{N} a_S  n(T)}{2}=\frac{\bar{N}}{N_0}S_{AL}[T], 
\label{change}
\ff
reflecting the fraction by which $\bar{N}$ differs from the  lapse of CDT.

In the continuum limit, we will have the same $g_c= \pm \frac{1}{2}$ and the change in (\ref{change})
just goes into rescaling of the renormalization group trajectory used to define the continuum limit, so that
the dimensionless $\lambda$ scales by $\lambda \rightarrow \bar{\lambda}= \lambda \bar{N}/N_0$.  

Quantities of physical interest, such as the Hausdorff dimension, are
computed by counting features of a typical triangulation, and taking
the continuum limit. For example, one counts the number of spatial
edges a fixed number of causal steps into the future and shows that 
this grows by the appropriate power. What is relevent for the 
continuum limits is the behavior of large sets, for a large number
of steps.  Allowing the lapses to vary does not change the physical
interpretation as long as the sets inolved are large enough that 
average quantities may be substituted for sums.  

It is important to emphasize that varying the lapse as proposed here
is not the same as summing over lapses.  What we have done amounts
to changing the gauge fixing constraint in the path integral, but
not removing it. The original causal dynamical triangulation model
amounts to a gauge fixed path integral, which is formally 
\f
Z= \int \prod_{x,t} \left(dq_{ab} dN dN^a
\delta (N-1 )\delta (N^a ) \Delta_{\rm FP}
\right) e^{i S[q_{ab},N ,N^a]} 
\ff
where $q_{ab}$ is the spatial part of the metric, $N$ is the lapse
and $N^a$ is the shift. What we have shown is that this may
be modified to 
\f
Z= \int \prod_{x,t} \left(dq_{ab} dN dN^a
\delta \left(N-f(x,t) \right)\delta (N^a ) \Delta_{\rm FP}
\right) e^{i S[q_{ab}, N ,N^a]} 
\ff
so long as, for all slices, 
\f
\frac{\int \sqrt {q} f(x,t)}{\int \sqrt {q}}=\mbox{constant}
\ff

Furthermore, substituting unit lapse with constant average within each
physical region involved in the measurement of a macroscopic
physical observable will not affect the physical quantities
that describe the macroscopic behavior such as the Hausdorff dimension
or the leading low energy behavior of the propagators of matter fields. 
We will discuss some examples of this in the next section.

\section{Matter coupling in $1+1$ dimensions}

We now consider the consequences of varying the lapse for models
where matter is coupled. Following \cite{c=1} we consider coupling
gravity to a spin system by adding a spin $\sigma_i$ to each triangle.  
The CDT coupled to gravity  partition function is 
\f
G(\Lambda , \beta, t) = \sum_{T(t)} e^{i S^{AL}[T ]}
\sum_{\sigma}e^{S^{ALspin} [T] },
\ff
where the initial and final lengths and spin states are fixed and
sum is over all the internal spins and histories with $t$ time steps. 
The matter action is given by
\f
S^{ALspin} [ T] = \beta \sum_{\langle ij\rangle\in  T} \sigma_i \sigma_j,
\label{matter1}
\ff
where $\beta$ is the coupling constant of the spin system. 

If the lapse varies, then the effect will be to modify (\ref{matter1})
to 
\f
S^{spin} [ T,  N ] = 
\sum_{\langle ij\rangle\in T }\sigma_i \sigma_t R_{ij} [N_i,N_j],
\ff
where the new factor $ R_{ij} [N_i,N_j]$  takes into account the varying metric.
However, note that, for each fixed triangulation and set of
lapses and each fixed set of spins,
\f
\sum_{\langle ij\rangle\in T} \sigma_i \sigma_j R_{ij} 
= \left ( \sum_{\{\langle ij\rangle: \sigma_i \sigma_j =1 \}} 1+ 
\sum_{\{\langle ij\rangle: \sigma_i \sigma_j =- 1 \}}-1 \right) R_{ij} 
=: P(\sigma ) \langle R_{ij} \rangle,
\ff
where $P(\sigma )$ is the number of nearest neighbor pairs
whose spins agree minus the number whose spins disagree. 

For the continuum limit we are interested in large triangulations,
so that the averages will be fixed to good accuracy over the sum
over triangulations. Hence, the effect of changing the gauge condition
so as to let the lapses vary can only be to modify the effective
matter coupling so 
\f
S^{spin} [T] = \beta^\prime \sum_{\langle ij\rangle\in T} \sigma_i \sigma_j,
\label{matter3}
\ff
where 
\f
\beta^\prime = \beta \langle R_{ij} \rangle.
\ff
Hence, the properties of matter found in \cite{c=1} for the continuum 
limit will be unchanged.

\section{CDT and locality}

We turn now to the question of whether the non-locality of the CDT model
is necessary for the existence of a good continuum limit.  Certainly, 
what is desired for a discete path integral approach to quantum
gravity is:
1) A good low energy or coarse grained limit, which is
both local and causal macroscopically, and leads to a recovery of
Lorentz invariant spacetime physics at low energies,
but also 
2) microscopic causality, and 
3) microscopic locality are desirable properties.
At present there are several models whose investigation is relevent for
the question of whether there exists a theory with all three good
properties.  It is interesting to compare the results concerning the
following models.

\begin{itemize}

\item{} CDT models\cite{AL}. As we discussed above, these are causal
microscopically, but
not local, because the evolution rule preserves the existentence of a global
slicing.  However, the evidence points to the existence of a good
continuum limit.

\item{} Causally evolving spin network models (CSN) \cite{CSN}. 
 In this construction
all edges are spacelike, and each triangle has a causal orienation.
There are two kinds of triangles: future pointing triangles with one
past edge and two future edges and past pointing triangles, with two
past edges and one future edge.
Histories are
constructed by a local rule which adds simplices locally.
The amplitude of a history is a product of the amplitude for each simplex added, and this is a function only of the simplices in the immediate past.  

\item{} Euclidean dynamical triangulation models (EDT) \cite{dynamical}
These
are constructed by a rule which is local, but not causal. One sums
over all $2d$ random triangulations, weighed only by the number of
triangles.  They do not lead to a continuum limit which reproduces
physics in relativistic spacetimes.

\item{} Causal set models \cite{cset}.  Histories are partial orders of events.  A generic causal set does not correspond to a triangulation of a manifold, nor is there an obvious notion of locality.
\footnote{Percolation causal set, in which the partial order is constucted by a percolation
process \cite{pcausalset} are microscopically causal and satisfy
a version of microscopic locality. }

\end{itemize}

There are other models, such as Regge calculus models and alternate
versions of causal spin foam models, but they are not needed for the
present discussion. We now discuss these, restricting to the $1+1$
or $2$ dimensional case, except in the causal set model, where there
is no microscopic notion of dimension.

We note that the causal set histories differ from the
histories in the other causal models, both CDT and CSN, in that the
antichains have no structure.  In both CDT and CSN the antichains are
spatial slices in a triangulation.

We next note that, since CDT does not allow topology change, there are 
many more EDT
histories
than CDT histories
(for given initial and final states,
specified by $l_0$ and $ l_t$, and given $N$).  The former do not have continuum
limits which reproduce physics which is macroscopically $1+1$
dimensional, while the latter do.   
Thus, the restriction to causal 
histories seems necessary for a good continuum limit.  A key question
is whether such a restriction must be non-local microscopically, as is
the case with CDT.   Since the continuum limit properties are known only for CDT and Euclidean histories, we will attempt a preliminary comparison of CSN with these two theories. 

The local CSN rules mean that it is possible to add many more simplices in some part of a spatial slice, while keeping other parts of it unchanged.  While this is not a topology change as such (it causes no branching of the spatial slice), such a CSN history can be mapped to a Euclidean triangulation of  a branching history.  (Note that there is no map of CSN histories to topologies with branching and rejoining spatial slices.)  In addition, since a CSN simplex also carries a causal structure, multiple CSN histories  (where the same triangulation carries different causal structures) can be mapped to the same Euclidean one.

There are then more CSN histories than Euclidean ones.  
This means that, if the CSN
histories are weighed equally by $e^{\imath \lambda N}$ as in CDT and EDT
histories, there can be no good continuum limit which reproduces
$1+1$ dimensional physics.  It has been argued (see for example \cite{LW}) that, since the growth in volume of the number of inequivalent triangulations of a given volume is factorial, no adjustments of the weights can stop models with topology change from having the continuum limit properties of the Euclidean dynamical triangulations.

Unfortunately, that is how far the implications of CDT for other path integral models can be taken.  The above argument is correct for a statistical path integral, namely only for EDT and the Wick rotated CDT.  For a quantum path integral, the possibility of interference and cancellations remains.  There is no Wick rotation for CSN, so the mapping of CSN histories to EDT ones does not imply anything conclusive about the CSN weights.

To summarize, at present, the only
of these models for which there is evidence that they reproduce $1+1$
dimensional physics in the low energy or continuum limit is the CDT model,
whose construction, however, appears to be non-local.  The other models are local
microscopically, but either they do not reproduce low-dimensional causal physics
macroscopically or there is no method to solve them.  
Thus, from the present results, it appears as if there is
a conflict between microscopic and macroscopic local physics as there is
no model that has both.  

\section{Conclusions}

We began by mentioning three features which distinguish the Abjorn-Loll causal dynamical triangulations model 
from other background independent approaches to the path integral in quantum gravity.  Of these, we have
seen that the restriction to unit lapse, stressed in point $2$ can be somewhat relaxed, providing support for the view that the preferred time slicings
of the model reflect a gauge fixing, rather than a physical preferred time.   

There are a number of further steps, which are currently under investigation. 

\begin{itemize}

\item{} We expect that the relaxation of the gauge fixing described here,
and in \cite{tomasz} for $2+1$ dimensions,  will hold also in the $3+1$ dimensional
model\cite{AL3+1}.   

\item{}The need for a constant time average lapse may depend on the setup of the transfer matrix
solution of the CDT.  To check this, an alternative ``spacetime'' solution of CDT in 1+1 is currently in progress \cite{AM}.

\item{} Given the recent results concerning the  CDT model in $3+1$ dimensions which support the conjecture that the model
has a good continuum limit which yields a theory of spacetime geometry in $3+1$ dimensions, 
it becomes of interest to see if that model can be
elaborated to yield a spin foam model with similar properties. This can be done by noting that spatial slicings exist
whose duals are graphs, and then extending the model by introducing labels as in spin foam models.  

\item{} It is of interest to understand better the relationship between causal spin foam models such as
those studied in \cite{CSN} and causal dynamical triangulation models.  This raises the issue of non-locality, 
as the latter's rules are non-local in ways the former's are not.  It is of particular interest to 
establish whether the good results gotten with the CDT models are dependent on
the presence of non-locality in the microscopic evolution rules.

\end{itemize}

\section*{ACKNOWLEDGEMENTS}

We are grateful to Jan Ambjorn, Renate Loll and members of Perimeter Institute for discussions on
these issues.


\begin{thebibliography}{99}

\bibitem{dynamical}J.~Ambjorn, Z.~Burda, J.~Jurkiewicz and C.~F.~Kristjansen,
``Quantum gravity represented as dynamical triangulations,''
Acta Phys.\ Polon.\ B {\bf 23}, 991 (1992); J.~Ambjorn,
``Quantum Gravity Represented As Dynamical Triangulations,''
Class.\ Quant.\ Grav.\  {\bf 12}, 2079 (1995);
M.~E.~Agishtein and A.~A.~Migdal,
``Simulations of four-dimensional simplicial quantum gravity,''
Mod.\ Phys.\ Lett.\ A {\bf 7}, 1039 (1992).

\bibitem{AL}J. Ambjorn, A. Dasgupta, J. Jurkiewicz and R. Loll,
``A Lorentzian cure for Euclidean troubles'', hep-th/0201104;
J. Ambjorn and R. Loll, Nucl. Phys. B536 (1998) 407 [hep-th/9805108];
J. Ambjorn, J. Jurkiewicz and R. Loll,
Phys. Rev. Lett. 85 (2000) 924 [hepth/ 0002050]; Nucl. Phys. B610 (2001) 347
[hep-th/0105267]; R. Loll, Nucl. Phys. B (Proc. Suppl.) 94
(2001) 96 [hep-th/0011194]; J. Ambjorn, J. Jurkiewicz and R. Loll, Phys.
Rev. D64 (2001) 044011 [hep-th/0011276];
J.~Ambjorn, J.~Jurkiewicz, R.~Loll and G.~Vernizzi,
{\it Lorentzian 3d gravity with wormholes via matrix models},
JHEP {\bf 0109}, 022 (2001)
[arXiv:hep-th/0106082]; 
B. Dittrich (AEI, Golm), R. Loll, {\it  A Hexagon Model for 3D Lorentzian Quantum 
Cosmology}, hep-th/0204210.

\bibitem{conformalresolve}A. Dasgupta and R. Loll, Nucl. Phys. B606
(2001) 357 [hep-th/0103186].

\bibitem{c=1}J. Ambjorn (Niels Bohr Institute), K.N. Anagnostopoulos (Univ.
of Crete), R. Loll, {\it Crossing the c=1 barrier in 2d Lorentzian quantum
gravity}, hep-lat/9909129, Phys.Rev. D61 (2000) 044010.

\bibitem{AL3+1}J. Ambjorn, J. Jurkiewicz and R. Loll, 
{\it Emergence of a 4D World from Causal Quantum Gravity}, 
 hep-th/0404156. 

\bibitem{CSN}Fotini Markopoulou, ``Dual formulation of spin network
evolution'', gr-qc/9704013;  
  {\it Quantum geometry with intrinsic local causality}, Fotini Markopoulou and Lee Smolin, gr-qc/9712067,  Phys. Rev. D 58
084032 (October 15, 1998), (1998).
 
\bibitem{cset}L.~Bombelli, J.~H.~Lee, D.~Meyer and R.~Sorkin,
``Space-Time As A Causal Set,''
Phys.\ Rev.\ Lett.\  {\bf 59}, 521 (1987)


\bibitem{pcausalset}X.~Martin, D.~O'Connor, D.~P.~Rideout and R.~D.~Sorkin,
``On the 'renormalization' transformations induced by cycles of expansion  and contraction
in causal set cosmology,''
Phys.\ Rev.\ D {\bf 63}, 084026 (2001)
[arXiv:gr-qc/0009063];
D.~P.~Rideout and R.~D.~Sorkin,
``Evidence for a continuum limit in causal set dynamics,''
Phys.\ Rev.\ D {\bf 63}, 104011 (2001)
[arXiv:gr-qc/0003117];  David Rideout,
{\it  Dynamics of Causal Sets}, 
 gr-qc/0212064; D. P. Rideout, R. D. Sorkin,
 {\it  A Classical Sequential Growth Dynamics for Causal Sets},
 gr-qc/9904062,  Phys.Rev. D61 (2000) 024002.  

\bibitem{LW} R Loll and W Westra, ``Space-time foam in 2D and the sum over topologies'', hep-th/0309012

\bibitem{AM}M Ansari and F Markopoulou, in preparation. 

\bibitem{tomasz}T. Konopka, {\it  Foliations and 2+1 Causal Dynamical Triangulation Models}, 
hep-th/0505004. 

\end{thebibliography}
\end{document}